\documentclass[a4paper]{amsart}
\usepackage{amsrefs}
\usepackage{xr-hyper}
\IfFileExists{eulervm.sty}{
\usepackage{eulervm}}{}

\usepackage[breaklinks=true,
colorlinks=true,
bookmarks=true,
hyperindex,
]{hyperref}
\usepackage{graphicx}
\let\cites=\cite

\providecommand{\citelist}{}
\providecommand{\amscite}[3]{\cite{#1}#2{#3}}
\usepackage{amssymb,amsthm,amsmath}
\usepackage{color}
\newtheorem{thm}{Theorem}
   \PassOptionsToPackage{british}{babel}
    \newtheorem{prop}[thm]{Proposition}

   \theoremstyle{definition}

    \newtheorem{defn}[thm]{Definition}

    \newtheorem{example}[thm]{Example}

   \theoremstyle{remark}

\providecommand{\norm}[2][\relax]{\left\|#2\right\|\ifx#1\relax\else_{#1}\fi}
\providecommand{\modulus}[2][\relax]{\left| #2 \right|\ifx#1\relax\else_{#1}\fi}
\providecommand{\oper}[1]{\mathcal{#1}}
\providecommand{\algebra}[1]{\ensuremath{\mathfrak{#1}}}
\providecommand{\Space}[3][]{\ifx#2R\ifx#1e \mathbb{C}^{#3} \else
\ifx#1p \mathbb{D}^{#3} \else
\ifx#1h \mathbb{O}^{#3} \else
\ifx#1\sigma \mathbb{A}\!^{#3} \else
\ensuremath{\mathbb{#2}^{#3}_{#1}{}} \fi \fi \fi \fi \else
\ensuremath{\mathbb{#2}^{#3}_{#1}{}} \fi}
\providecommand{\FSpace}[3][]{\ensuremath{\ifx#2l \ell_{#3}^{#1}{}\else
  \mathsf{#2}_{#3}^{#1}{}\fi}}

\providecommand{\uir}[3][0]{\ifcase #1{\rho^{#2}_{#3}}% 
\or {\breve{\rho}^{#2}_{#3}}%
\or {\tilde{\rho}^{#2}_{#3}}\fi}
\usepackage{braket}
\providecommand{\scalar}[2]{\braket{#2|#1}}
\providecommand{\rmi}{\mathrm{i}}
\providecommand{\rme}{\mathrm{e}}
\providecommand{\rmd}{\mathrm{d}}

\providecommand{\myhbar}{\hslash}
\providecommand{\myh}{h}
\providecommand{\map}[1]{\mathsf{#1}}

\providecommand{\Zbl}[1]{Zbl\href{http://www.emis.de:80/cgi-bin/zmen/ZMATH/en/zmathf.html?first=1&maxdocs=3&type=html&an=#1&format=complete}{#1}}

\providecommand{\myeprint}[2]{\href{#1}{\texttt{#2}}}
\DeclareFontFamily{OT1}{cyr}{}
\DeclareFontShape{OT1}{cyr}{m}{n}
   {  <5> <6> <7> <8> <9> gen * wncyr
      <10> <10.95> <12> <14.4> <17.28> <20.74> <24.88> wncyr10}{}
\DeclareFontShape{OT1}{cyr}{m}{it}
    {
       <5> <6> <7> <8> <9> gen * wncyi
      <10> <10.95> <12> <14.4> <17.28> <20.74> <24.88>wncyi10
      }{}
\DeclareFontShape{OT1}{cyr}{m}{ss}
    {
       <5> <6> <7> <8> wncyss8
       <9> wncy9
      <10> <10.95> <12> <14.4> <17.28> <20.74> <24.88>wncyss10
      }{}
\DeclareFontShape{OT1}{cyr}{m}{sc}
    {
       <5> <6> <7> <8> <9> <10> <10.95> <12> <14.4> <17.28> <20.74> <24.88>wncysc10
      }{}
\DeclareFontShape{OT1}{cyr}{bx}{n}
   {
       <5> <6> <7> <8> <9> gen * wncyb
      <10> <10.95> <12> <14.4> <17.28> <20.74> <24.88>wncyb10
      }{}
\DeclareTextFontCommand{\textcyr}{\fontfamily{cyr}\selectfont}
\providecommand{\cyr}{\fontfamily{cyr}\selectfont\def\cprime{\~}}
\providecommand{\cprime}{'}

\begin{document}
\title[Schr\"odinger Equation and a Coherent States Transform]%
{Solving the Schr\"odinger Equation by Reduction to a
  First-order Differential Operator through a Coherent States
  Transform}
%\shorttitle{Schr\"odinger Equation and a Coherent States Transform}
\date{\today}

\author{Fadhel Almalki}
\email{fadhel1408@gmail.com}
% %\homepage[]{Your web page}
\thanks{Sponsored by The Taif University  (Saudi Arabia).}
% \altaffiliation{}

\author{Vladimir V. Kisil}
%\homepage%[]
\urladdr
{http://www.maths.leeds.ac.uk/~kisilv/}
\email%[]
{V.Kisil@leeds.ac.uk}
%{kisilv@maths.leeds.ac.uk}
%\affiliation
\address
{School of Mathematics,
  University of Leeds,
  Leeds LS2\,9JT,
  UK}

%\author{Fadhel Almalki\thanks{Sponsored by The Taif University  (Saudi Arabia).} and Vladimir V. Kisil}
%\shortauthor{F. Almalki and V.V. Kisil}

% \institute{School of Mathematics,
%  University of Leeds,
%  Leeds LS2\,9JT,
%  UK}

%\pacs{03.65.-w}{Quantum mechanics}
%\pacs{02.30.Jr}{Partial differential equations}
%\pacs{03.65.Db}{Functional analytical methods}
\begin{abstract}
  The Legendre transform expresses dynamics of a classical system through
  first-order Hamiltonian equations. We consider coherent state
  transforms with a similar effect in quantum mechanics: they reduce
  certain quantum Hamiltonians to first-order partial differential
  operators. Therefore, the respective dynamics can be explicitly solved through
  a flow of points in extensions of the phase space. This
  generalises the geometric dynamics of a harmonic oscillator in the
  Fock space. We describe all Hamiltonians which are geometrised
  (in the above sense) by Gaussian and Airy beams and write down
  explicit solutions for such systems.
\end{abstract}
% \pacs
%%AMSMSC:

%\begin{document}
\maketitle

\section{Introduction}
\label{sec:introduction}

The coherent states (CS) were introduced by Schr\"odinger in
1926~\cite{Schrodinger26a} but were not in use until much
later~\cites{Bargmann61,Segal60,Glauber63a,Sudarshan63a}. Nevertheless,
ideas from~\cite{Schrodinger26a} were crucial for formation of quantum
mechanics~\cite{Steiner88a}. Naturally, further developments of the
concept of CS manifested a remarkable depth and
width~\cites{KlaSkag85,Perelomov86,Walls08,AliAntGaz14a,Gazeau09a}.
The canonical CS of a harmonic oscillator have a variety of important
properties, e.g. semi-classical dynamics, minimal uncertainty,
parametrisation by points of the phase space, resolution of the
identity, covariance under a group action, etc. As usual, different
generalisations of CS start from one particular property used as a
definition, other properties may or may not follow as consequences
depending on circumstances. For example, the approach in~\cite{Hall94a} employed a connection of CS with the heat kernel (Gaussian) and its analytical extension, but requires the compactness of the underlying group---in contract to the original setup of CS from the Heisenberg group.

This letter revises geometrisation of quantum evolution in CS
representation. More specifically, in Defn.~\ref{de:method-order-reduction} we \emph{describe a method} to determine all Hamiltonians admitting a reduction to a first order partial differential equation (PDE) by means of the coherent state transform  (CST) with a given fiducial vector. Thereafter, the corresponding first order PDE can be explicitly solved and the geometrised evolution is realised by a time-dependent coordinate transformations. The method is based in the development~\cites{%
  Kisil11c,%
  Kisil13c,%
  Kisil17a,%
AlmalkiKisil18a} of the CS from group representations~\cites{Perelomov86,AliAntGaz14a,Gazeau09a}, notably an analyticity-type condition in the image space of CST as explained below. 

Our inspiration is
that the canonical CS~\cites{Schrodinger26a,Glauber63a,Sudarshan63a,%
  Bargmann61,Segal60,Walls08,Gazeau09a}
are parametrised by points \(z=q+\rmi p\) of the phase space in such a
way that the time evolution of a quantum state \(\tilde{f}(t,z)\) in a
harmonic potential is given by
\begin{equation}
  \label{eq:harmonic-oscillator-dynamics}
  \tilde{f}(t,z)= e^{-\pi \rmi \omega t} f(0, e^{-2\pi \rmi \omega t} z)\,.
\end{equation}
Its key ingredient is the rigid rotations
\(z\mapsto \rme^{-2\pi\rmi \omega t} z\) of the phase space---the
dynamics of a classical harmonic
oscillator. Thus, the geometrisation of the evolution also manifests the correspondence principle between quantum and classical mechanics.
% The correspondence principle is manifested because the same map
% describes the classical harmonic oscillator evolution. However, the
% distinction of quantum and classical mechanics reflected in numerous
% no-go theorems does not allow to have exactly the same realisation
% of the correspondence principle for all Hamiltonians. Thus, we 
% need different CS adopted for particular Hamiltonians as specified in
% the following definition.

However, various ``no-go'' theorems suggest that this correspondence
cannot be universal.  Therefore,
there are many different approaches to geometrisation of quantum
dynamics~\cites{%
  Kibble79a,%,
  Prugovecki82a,%
  BrookePrugovecki85a,%
  Zachos02a,%
  CarienaClemente-GallardoMarmo07a,%
  ShalashilinBurghardt08,%
  KaramatskouKleinert14a,%
  Tavernelli16a,%
  HerranzDeLucasTobolski17a,%
  CiagliaDiCosmoIbortMarmo17a%
}, see also~\cite{HurleyVandyck09a,Clemente-GallardoMarmo08a} for
surveys and further references.  Those works are often motivated by
conceptual considerations of fundamental structures of Lagrangian and Hamiltonian dynamics. Our main aim here
is more pragmatic: we look for a simple and effective method to
express a quantum evolution through a flow of points of some set. Few
connections to the above papers will be discussed at the end of this letter.

Let the dynamic of a quantum system be defined by a Hamiltonian \(H\) and
the respective Schr\"odinger equation
\begin{equation}
  \label{eq:schtodinger-equation}
  \rmi\myhbar \dot{\phi} (t) =H \phi (t).
\end{equation}
Geometrisation of~\eqref{eq:schtodinger-equation} suggested
in~\cite{CiagliaDiCosmoIbortMarmo17a} uses a collection
\(\ket{x}\) of CS parametrised by points of a set
\(X\). Then the solution \(\ket{x,t}\) of~\eqref{eq:schtodinger-equation}
for an initial value \(\ket{x,0}=\ket{x}\) shall have the form
\begin{equation}
  \label{eq:geometrization-solution}
  \ket{x, t}=\ket{x_t}, 
\end{equation}
where \(x_t\) is an orbit of a one-parameter group of
transformations \(X \rightarrow X\). Recall that the coherent state
transform  (CST) \(\tilde{f}(x)\) of a state \(f\) is defined by
\begin{equation}
  \label{eq:CST-defn}
  f \mapsto \tilde{f}(x)=\scalar{f}{x}.
\end{equation}
It is common that CST is a unitary map onto a subspace
\(\FSpace{F}{2}\) of \(\FSpace{L}{2}(X,d\mu)\) for a suitable measure
\(d\mu\) on \(X\). If CS \(\ket{x}\), \({x\in X}\) geometrise a
Hamiltonian \(H\) in the above sense, then for CST
\(\tilde{f}(t,x)=\scalar{f(t)}{x}\) of an arbitrary solution
\(f(t)=\rme^{-\rmi t H/\myhbar} f(0)\)
of~\eqref{eq:schtodinger-equation} we have:
\begin{align*}
  \tilde{f}(t,x)&=\scalar{\rme^{-\rmi t H /\myhbar } f(0)}{x}%\\
  % &
      =\scalar{f(0)}{x_t}
  =\tilde{f}(0,x_t).
\end{align*}
Thus, if CS geometrise a Hamiltonian \(H\) then the dynamic of any
image \(\tilde{f}\) of the respective CST is given by a transformation
of variables.

It was already noted in~\cite{CiagliaDiCosmoIbortMarmo17a} that even
the archetypal canonical CS do not geometrise the harmonic oscillator
dynamics in the above strict sense due to the presence of the overall
phase factor in the
solution~\eqref{eq:harmonic-oscillator-dynamics}. The factor is not a
minor nuisance but rather a fundamental element: it is responsible for
a positive energy of the ground state.

To accommodate such observation with geometrising, we propose the
adjusted meaning of geometrisation:
\begin{defn}
  \label{de:geometrizing-CS}
  A complete collection \(\ket{x}\) of CS parametrised by points of a
  manifold \(X\) geometrises a quantum dynamic, if the time evolution
  of the CST \(\tilde{f}=\scalar{f}{x}\) is defined by a Schr\"odinger equation
  \begin{equation}
    \label{eq:schtodinger-equation-geom}
    \rmi\myhbar {\frac {\rmd  \tilde{f}}{\rmd t}}  = \tilde{H} \tilde{f}\,,
  \end{equation}
  where \emph{\(\tilde{H}\) is a first-order differential operator} on
  \(X\). 
\end{defn}
%If   \(X\) is a real manifold then \(\tilde{H}\) has purely imaginary
%coefficients. 

To find geometrising CS one can use symmetries of the Schr\"odinger
equation and a Hamiltonian \(H\), cf.~\cites{Niederer72a,%
  KalninsMiller74a,%
  ATorre09a,%
  Niederer73a,%
  Wolf76a,%
  AldayaGuerrero01a,%
  AldayaCossioGuerreroLopez-Ruiz11b,%
  AldayaGuerreroMarmo98a}. In
particular~\cites{Perelomov86,AliAntGaz14a,Gazeau09a}, group
representations are a rich source of CS. More precisely, let \(X\) be the
homogeneous space \(G/H\) for a group \(G\) and its closed subgroup
\(H\). Then, for a representation \(\uir{}{}\) of \(G\) in a space
\(V\) and a fiducial vector \(\ket{0}\in V\) the collection of coherent
states is defined by
\begin{equation}%{displaymath}
  \ket{x}=\uir{}{}(\map{s}(x)) \ket{0}, 
\end{equation}%{displaymath}
where \(x\in G/H\) and \(\map{s}: G/H \rightarrow G\) is a Borel section. 
\begin{example}
  \label{ex:canonical-CS}
  The canonical CS~\cites{Schrodinger26a,Glauber63a,Sudarshan63a,%
    Bargmann61,Segal60,Walls08,Gazeau09a} are produced by \(G\) being
  the Heisenberg
  group~\cites{Folland89,Kirillov04a,Kisil02e,Kisil17a}, \(H\)---the
  centre of \(G\), \(\uir{}{}\)---the Schr\"odinger representation and
  \(\ket{0}\) be Gaussian \(\rme^{-\pi \myhbar m\omega x^2}\). These
  CS geometrise the harmonic oscillator dynamic with the specific
  potential in the Hamiltonian:
  \(H=\frac{1}{2}\left(\frac{1}{m}p^2+m \omega^2 q^2\right)\). Indeed,
  the corresponding CST maps states to the Fock--Segal--Bargmann space
  and the harmonic oscillator
  dynamics~\eqref{eq:harmonic-oscillator-dynamics} satisfies the first
  order differential equation:
  \begin{equation}
    \rmi\myhbar {\frac {\rmd}{\rmd t}} \tilde{f}(t,z)=
    \pi\myhbar\omega (\tilde{f}(t,z) + 2 z\partial_z\tilde{f}(t,z)),
    \quad z\in\Space{C}{}.
  \end{equation}
\end{example}
The construction of CS from the group representations is fully
determined by a choice of \(G\), \(H\), \(\uir{}{}\) and \(\ket{0}\).
Thus, varying some of these components we obtain different
geometrisable Hamiltonians in sense of Defn.~\ref{de:geometrizing-CS}.
For example, the minimal nilpotent extension of the Heisenberg group
\(\Space{G}{}\) (see~\eqref{anh comutation relation}) allows to use
Gaussian \(\rme^{-\pi \myhbar E x^2}\) with arbitrary squeeze
parameter \(E\)~\cites{Walls08,Gazeau09a} as a fiducial vector
\(\ket{0}\) for simultaneous geometrisation of all harmonic
oscillators with different values of
\(m\omega\)~\citelist{\cite{AlmalkiKisil18a} \cite{Child14a}*{\S~8.2}}.

\begin{defn}
  \label{de:method-order-reduction}
The \emph{method of order reduction} employed in this letter consists of the following steps. For a group \(G\) and its unitary irreducible representation \(\uir{}{}\):  
\begin{enumerate}
\item Chose a fiducial vector \(\ket{0}\) which is annihilated by certain form of \(\uir{}{}\), cf.~\eqref{eq:derived-representation-first};
\item Calculate~\cites{%
  Kisil11c,%
  Kisil13c,%
  Kisil17a,%
  AlmalkiKisil18a} the respective operator \(\mathcal{C}\), cf.~\eqref{analyticity operator airy}, which annihilates the image space of the CST~\eqref{eq:CST-defn}.  
\item Find all Hamiltonians of the form \(H = F + A \mathcal{C}\), where  \(F\) is a first order PDE and \(A\) is any operator.
\end{enumerate}
\end{defn}

The above method is quite general and is not limited to a particular group or representation. As an illustration, we use it in the following extension of the classic setup from Ex.~\ref{ex:canonical-CS}:
\begin{itemize}
\item The Heisenberg group is extended to the minimal nilpotent step 3 group
  \(\Space{G}{}\) defined in~\eqref{anh comutation
    relation}. Consequently our CS are parametrised by an extension of
  the classical phase space. 
\item The fiducial vector \(\ket{0}\) is changed from the Gaussian to
  a cubic exponent~\eqref{mother-wavelet for kirilov re},
  cf. Fig.~\ref{fig:graund-state-harm}. The later is the Fourier
  transform of an Airy wave packet~\cite{BerryBalazs79a} which are
  useful in paraxial optics~\cite{ATorre09a,ATorre14a}.
\end{itemize}

For the found geometrisable Hamiltonians we
can write explicit generic solutions through well-known integral
transforms.  The letter is concluded by a discussion a wider framework for our method and some its further usage is outlined.
 
\section{Background}
\label{sec:background}
Here we briefly introduce the necessary background. It was already
used in~\cite{AlmalkiKisil18a}, which contains a detailed presentation
and further references.

% \subsection{The shear group $\Space{G}{}$}
% \label{sec:shear-group-A}
Let \(\algebra{g}\) be the minimal nilpotent step \(3\) Lie algebra
spanned by \(\{X_1,X_2,X_3,X_4\}\) with the only non-vanishing
commutators~\cites{CorwinGreenleaf90a,%
  Kirillov04a,%
  AlmalkiKisil18a,%
  AllenAnastassiouKlink97,%
  JorgensenKlink85,%
  Klink94,%
  BacryLevy-Leblond68a,%
GhaaniFarashahi17a}:
\begin{equation}
  \label{anh comutation relation}
  [X_1,X_2]=X_3,\qquad [X_1,X_3]=X_4\,.
\end{equation}
The corresponding Lie group \(\Space{G}{} \sim \Space{R}{4}\) is three-step nilpotent 
 and its elements will be
denoted by
\begin{equation}%{displaymath}
  g=(x_1,x_2,x_3,x_4):=\rme^{x_4X_4}\rme^{x_3X_3}\rme^{x_2X_2}\rme^{x_1X_1}\,.
\end{equation}%{displaymath}
The group can be physically interpreted as a central extension by
\(X_4\) of the Galilean group spanned by \( X_1\), \(X_2\) and
\(X_3\), see~\cite{AlmalkiKisil18a} for the related discussion. The r\^ole of central extension in quantisation is described, for
example, in~\cite{AldayaGuerreroMarmo98a}.

There are two important subgroups of  \(\Space{G}{}\): the centre \(Z\)
and the subgroups of elements \((x_1,0,x_3,x_4)\), which is isomorphic
to the Heisenberg group. On the other hand \(\Space{G}{}\) is a
subgroup of the Schr\"odinger group~\cites{Niederer72a,%
  KalninsMiller74a,%
  ATorre09a,%
  Niederer73a,%
  Wolf76a,%
  AldayaGuerrero01a,%
  AldayaCossioGuerreroLopez-Ruiz11b%
}.

For two real parameters \(\myh_2\) and  \(\myhbar_4\neq 0\), we will use the unitary irreducible (UIR) representation of the group \(\Space{G}{}\) in
\(\FSpace{L}{2}(\Space{R}{})\) given by, cf.~\citelist{\amscite{Kirillov04a}*{\S\,3.3, (19)} \cite{AlmalkiKisil18a}}:
\begin{equation}
  \label{irreducible repre of A}
  [\uir{}{\myh_2 \myhbar_4}(g)f](x'_1)
  =\rme^{ 2\pi\rmi(\myh_2x_2+   \myhbar_4(x_4-x_3x_1'+\frac{1}{2}x_2x_1'^2))}f(x'_1-x_1).
\end{equation}
Note that \(\uir{}{\myh_2 \myhbar_4}(x_1,0,x_3,x_4)\) coincides with
the Schr\"odinger representation of the Heisenberg
group~\cite{Kisil02e,Kisil10a,Kisil17a}. In particular, vectors
\(X_1\) and \(X_3 \in \algebra{g}\) correspond to momentum and
position observables in the coordinate representation.

% \subsection{The induced coherent state transform}
% \label{sec:induc-wavel-transf}
Let \(\uir{}{}\) be a representation of  \(G\) and \(\ket{0}\) be
a joint eigenvector of operators \(\uir{}{}(h)\) for all \(h\) in
subgroup \(H\) of \(G\):
\begin{equation}
  \label{H induced wavelet proprety}
  \uir{}{}(h)\ket{0}=\chi(h) \ket{0} \quad \text{for all} \quad h\in H,
\end{equation}
where \(\chi\) is a character of \(H\). Then, CST is
completely determined by its values on \(G/H\)~\cites{Gilmore1972a,Perelomov86}. Thus, for a section \(\map{s}: G/H \rightarrow G\) % , which is a right inverse to the
% natural projection \(G\rightarrow G/H\),
we define the \emph{induced coherent state transform} (ICST)
from a Hilbert space \(\FSpace{H}{}\) to a space of
functions \(\FSpace{L}{0}(G/H)\) by:
 \begin{equation}
 \label{induced wavelet transform}
\tilde{f}(x)=\scalar{f}{x}, \text{ where }
\ket{x}=\uir{}{}(\mathsf{s}(x))\ket{0}, \ x\in G/H\,.
 \end{equation}

For the subgroup \(H\) being the centre
\(Z% =\{(0,0,0,x_4)\in\Space{A}{}: x_4\in \Space{R}{}\}
\) of
\(\Space{G}{}\), the representation
\(\uir{}{\myh_2\myhbar_4}\)~\eqref{irreducible repre of A} and the
character \(\chi(0,0,0,x_4)=\rme^{2\pi\rmi \myhbar_4 x_4}\) any
function in \(\FSpace{L}{2}(\Space{R}{})\) satisfies the
eigenvector property~\eqref{H induced wavelet proprety}. Thus, for
the respective homogeneous space \(\Space{G}{}/Z\sim \Space{R}{3}\)
and the section
\(\mathsf{s}:\Space{G}{}/Z\to \Space{G}{}; \
\mathsf{s}(x_{1},x_{2},x_{3})=(x_{1},x_{2},x_{3},0)\),  
 ICST is:
\begin{align}
  \label{eq:wavelet-transform-shear-group}
  &\tilde{f}(x_{1},x_{2},x_{3})= \scalar{f}{x_{1},x_{2},x_{3}}\\
  % \nonumber
  % &=\scalar{ f}{x_{1},x_{2},x_{3}}\\
  % \nonumber
  % &=\int_{\Space{R}{}} f(y)\, \overline{\uir{}{\myh_2\myhbar_4}(x_{1},x_{2},x_{3},0)\,\phi(y)}\,\rmd y
  % \\
  % \nonumber
  % &=\int_{\Space{R}{}} f(y)\,\rme^{-2\pi \rmi (\myh_2x_2+\myhbar_4(-x_3y+\frac{1}{2}x_2y^2))}\,
  % \overline{\phi}(y-x_1)\,\rmd y\\
  \nonumber
  &=\rme^{-2\pi \rmi \myh_2x_2}\int_{\Space{R}{}} \rmd y \,
    f(y)\,\rme^{-2\pi \rmi \myhbar_4(-x_3y+\frac{1}{2}x_2y^2)}\,
                                         \overline{\phi}(y-x_1).
\end{align}
It intertwines \(\uir{}{\myh_2\myhbar_4}\) with the representation
%of \(\Space{G}{}\)
on \(\FSpace{L}{2}(\Space{R}{3})\):
\begin{align}
  \nonumber
  &[\uir[2]{}{\myhbar_4}(g)f](x_1',x_2',x_3')
=\rme^{2\pi \rmi \myhbar_4(x_4-x_1x_3+\frac{1}{2}x_1^2x_2+x_1x_3'-\frac{1}{2}x_1^2 x_2')}\\
  \label{reducible re of anhar def}
  &\qquad \times f(x_1'-x_1,x_2'-x_2,x_3'-x_3-x_1x_2'+x_1x_2).
\end{align}
For a fixed \(x_2 \in \Space{R}{}\), the map
\(f \otimes \ket{0} \mapsto \tilde{f}(\cdot,x_2,\cdot)\)
is a unitary operator: \(\FSpace{L}{2}(\Space{R}{})\otimes
\FSpace{L}{2}(\Space{R}{}) \rightarrow \FSpace{L}{2}(\Space{R}{2})\).
% \subsection{Operators annihilating the space
%   ICST image space}
\begin{figure}[htbp]
  \centering
  \includegraphics{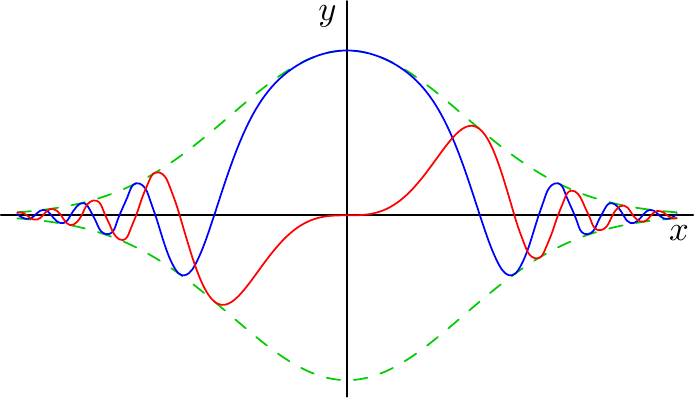}  
  \caption[Ground state of harmonic oscillator in a field]
  {Fiducial vector \(\phi\)~\eqref{mother-wavelet for kirilov re}:
    solid blue and red graphs are its real and imaginary parts
    respectively. The green dashed envelope is the absolute value
    \(\modulus{\phi}\), which coincides with a Gaussian.}
  \label{fig:graund-state-harm}
\end{figure}
Define the cubic extension  \(\ket{0}\) of a Gaussian, see
Fig.~\ref{fig:graund-state-harm} by:
\begin{equation}
  \label{mother-wavelet for kirilov re}
  \phi(y)=\exp\left(\frac{ \pi\rmi D\myhbar_{4} }{3} y^{3}-  \pi E\myhbar_{4}y^{2}+ 2\pi\rmi D\myh_{2}y\right),
\end{equation}
where square--integrability of \(\ket{0}\) over \(\Space{R}{}\) requires
that \(E\myhbar_4\) is strictly positive and \(D\) is real.
Physically,  the parameter \(E\) encodes a squeeze of CS~\cite{AlmalkiKisil18a}.
State \(\ket{0}\)~\eqref{mother-wavelet for kirilov re} is 
interesting because it is a null-solution of the
generic derived representation \(\rmd\uir{}{\myh_2\myhbar_4}\) of
\(\algebra{g}\):
\begin{equation}
  \label{eq:derived-representation-first}
  \begin{split}
    % \lefteqn
    {\rmd\uir{\rmi X_{1}+\rmi DX_2 + E X_3}{\myh_2\myhbar_4}
      = \rmi
      \rmd\uir{X_{1}}{\myh_2\myhbar_4}
      +\rmi D
      \,\rmd\uir{X_{2}}{\myh_2\myhbar_4}
      +E\,\rmd\uir{X_{3}}{\myh_2\myhbar_4}}&\\
    \qquad =-\rmi \frac{\rmd\ }{\rmd y}- \pi   \myhbar_4
    D y^2-2\pi\rmi E\myhbar_4 y  -2\pi\myh_2 D& 
    \,.
  \end{split}
\end{equation}
Thus, this operator is a counterpart of the annihilation operator of
canonical CS. It follows~\cites{%
  Kisil11c,%
  Kisil13c,%
  Kisil17a,%
AlmalkiKisil18a}, the image space
\(\FSpace{L}{0}(\mathbb{G}/Z)\) of
ICST~\eqref{eq:wavelet-transform-shear-group} with the fiducial
vector~\eqref{mother-wavelet for kirilov re} satisfy to
\begin{equation}
\label{eq:analiticity-airy}
\mathcal{C}f(x_1,x_2,x_3)=0,\quad \text{for any}\,\, f\in \FSpace{L}{0}(\mathbb{G}/Z)
\end{equation}
where
\begin{align}
  \label{analyticity operator airy}
  \mathcal{C}% &= -\rmi \linv{X_1}-\rmi D\linv{X_2}+E\linv{X_3}\\
  % \nonumber 
             &=-\rmi \partial_{1}-\rmi  D\partial_{2} + (E -\rmi  D x_1)\partial_{3} -\pi \myhbar_4(2 \rmi   E  x_1+ D  x_1^{2})I\,,
\end{align}
which we call the \emph{analytic condition}.

Another condition on \(f\in \FSpace{L}{0}(\mathbb{G}/Z)\) is
generated by the Casimir element
\(C=X_3^{2}-2X_2X_4\)~\cites{CorwinGreenleaf90a,%
Kirillov04a}. The corresponding operator acts
as a multiplication operator by \(8\pi^2\myh_2\myhbar_4\) on
\(\FSpace{L}{0}(\mathbb{G}/Z)\), thus, for any \(f\in\FSpace{L}{0}(\mathbb{G}/Z)\) we have  \(
\oper{S}f(x_1,x_2,x_3)=0,\) where
\begin{align}
  \label{eq:general-annih}
  \oper{S}&=(\rmd\uir[2]{X_3}{\myhbar_4} )^2-2\,
            \rmd\uir[2]{X_3}{\myhbar_2}\, \rmd\uir[2]{X_4}{\myhbar_4}  - {8 \pi^2\myh_2\myhbar_4} I\\
  \nonumber
          &=  \partial_{33}^{2}+{4\pi \rmi}  \myhbar_4 \partial_{2}-8\pi^2  \myh_2  \myhbar_4 I\,.
\end{align}
The relation~\eqref{eq:general-annih} will be called the
\emph{structural condition} determined by the Casimir operator and  is
independent from a fiducial vector being used. Summing up, we conclude that
\emph{\(\FSpace{L}{0}(\mathbb{G}/Z)\) is annihilated by elements of
  the left operator ideal \(K\) generated by \(\oper{C}\)~\eqref{analyticity
    operator airy} and \(\oper{S}\)~\eqref{eq:general-annih}}.

\section{Geometrisation %of dynamics
  with Fourier--Fresnel and Fourier--Airy transforms}

Recall, any two operators different by an element of the left operator
ideal \(K\) generated by \(\oper{C}\)~\eqref{analyticity operator
  airy} and \(\oper{S}\)~\eqref{eq:general-annih} have equal
restrictions to the space \(\FSpace{L}{0}(\mathbb{G}/Z)\). Among
many equivalent operators we can look for a representative with
desired properties, e.g. a first order differential operator, which
geometrises dynamics in the sense discussed above.

In this section we present the complete characterisation of quadratic
forms on the Lie algebra \(\algebra{g}\) 
% \eqref{anh comutation relation}
 which admit geometrised
dynamics by means of covariant transform with a fiducial
vector~\eqref{mother-wavelet for kirilov re}.

% \subsection{Hamiltonians admitting first-order
% reduction}
% \label{sec:class-hamilt-admitt}

Let \(\uir{}{}\) be UIR of 
\(\Space{G}{}\), for the respective derived representation
\(\rmd \uir{}{}\) of \(\algebra{g}\) consider a general quadratic
form:
\begin{align}
  \label{general Hamiltonian on g}
  \rmd\uir{G}{}=\sum_{j,k=1}^{3} a_{jk}\rmd\uir{X_j}{} \rmd\uir{X_k}{} \qquad \text{where } X_n \in \algebra{g}\,.
\end{align}
If \(\rmd\uir{G}{}\) admits a geometric dynamic in Airy-type CS
\(\ket{x_1,x_2,x_3}\)
% parametrised by \((x_1,x_2,x_3,0)\in \Space{G}{}/Z\),
then there is a first-order differential operator,
\(H_r\) on \(\FSpace{L}{0}(\Space{G}{}/Z)\) such that
\(H_r-\rmd\uir[2]{G}{}\) is in the ideal generated by
\(\oper{C}\)~\eqref{analyticity operator airy} and
\(\oper{S}\)~\eqref{eq:general-annih}:
\begin{equation}
  \label{reduced Hamiltonian equation}
  H_r  = \rmd\uir[2]{G}{}+(A\partial_1+B\partial_2+C\partial_3+K)\mathcal{C}
  + F\mathcal{S}\,. 
\end{equation} 
Here \(A,B,C,K\) and \(F\) are certain coefficients, which are
chosen to eliminate all possible second-order derivatives 
in \(H=\rmd\uir[2]{G}{}\).
In its turn, this depends on values \(D\) and \(E\) in the fiducial
vector \(\ket{0}\)~\eqref{mother-wavelet for kirilov re} and the
respective analyticity operator \(\oper{C}\)~\eqref{analyticity operator airy}.

 Substituting the explicit
  expressions of the
  derived representation and \(\oper{C}\)~\eqref{analyticity operator airy},
 \(\oper{S}\)~\eqref{eq:general-annih}  into the relation
 \eqref{reduced Hamiltonian equation} we can eliminate second
  derivatives in \(H_r\) if:
\begin{align}
  \nonumber 
    A&=-\rmi  a_{11},\qquad
       B =\rmi  D a_{11}-\rmi( a_{21}+a_{12})\,,\\
  \nonumber 
  C&=-a_{11}(2 \rmi x_2 -\rmi  D x_1+ E)-\rmi (a_{13}+a_{31})\,,\\
  \nonumber 
    F&= -a_{11}u_{2}^2+(a_{31}+a_{13})u_{2}-a_{33}\,, \quad
       [u_2=Dx_1-x_2+\rmi E]\\
  \nonumber 
    K&=2 \pi \myhbar_4(a_{13}+a_{31})  x_1-a_{11}\big(-2\pi\rmi \myhbar_4 E x_1 +\frac{D}{E}\\
    &\qquad +5\pi \myhbar_4 D  x_1^{2} -4\pi \myhbar_4 x_1 x_2\big)+\frac{a_{21}}{E}\,.
\label{reduction polynomials}
\end{align}
The last parameter \(K\) is used to get imaginary coefficients in
front of \( \partial_{1}\) and \( \partial_{3}\) to obtain a geometric
action in the phase space parametrised by momentum-position variables
\((x_1,x_3)\). Additionally to the above values~\eqref{reduction
  polynomials},  we have to apply the following restriction of the
coefficients \(a_{jk}\) on the quadratic form~\eqref{general
  Hamiltonian on g}:
\begin{equation}
\label{Hamiltonian coefficients}
\begin{split}
  a_{12}&=2  D a_{11}-a_{21},\qquad
  a_{22}= D^{2} a_{11},\\
  a_{23}&=  D (a_{13} + a_{31})-a_{32}\,.
\end{split}
\end{equation}
The remaining coefficients \(a_{11}, a_{21}, a_{13}, a_{31},a_{32}\)
and \(a_{33}\) as well as parameters \(D\) and \(E\) are free
variables.
%\end{example}
Thus, we  have obtained the
desired classification:
\begin{prop}
  \label{pr:classification}
  The Hamiltonian~\eqref{general Hamiltonian on g} can be geometrised
  over \(\Space{G}{}/Z\) by Airy-type CS from the fiducial
  vector~\eqref{mother-wavelet for kirilov re} if and only if
  coefficients \(a_{jk}\) satisfy~\eqref{Hamiltonian coefficients}.
\end{prop}
Note that even the case \(D=0\) (that is a Gaussian as a fiducial
vector) is more general than the classical result of Schr\"odinger
from Example~\ref{ex:canonical-CS}: the latter requires the exact
match of the squeezing parameter \(E=m\omega\) in the Hamiltonian and
the CS. The larger group \(\Space{G}{}\) can
treat a harmonic oscillator through Gaussians with arbitrary squeeze, see~\cite{AlmalkiKisil18a} or presentation without groups in~\cite{Child14a}*{\S~8.2}.

\section{Solving the geometrised %Schr\"odinger
  equation}

Geometrisable Hamiltonians described in Prop.~\ref{pr:classification} can be
explicitly solved using the following steps:
\begin{enumerate}
\item Find the generic
  solutions of the analytic condition~\eqref{analyticity operator airy},
  which is a first-order PDE. Since it is Hamiltonian-independent,
  the obtained solution can be re-cycled.
\item Substitute the above analytic solution into the reduced
  (first-order) form of the Schr\"odinger equation for \(H_r\)~\eqref{reduced Hamiltonian equation}. A resulting
  dynamical equation is significantly simplified in analytic
  variables and admits an explicit generic solution.
\item Substitute the generic solution into the structural
  condition~\eqref{eq:general-annih}, this produces a second-order
  PDE solved by well-known integral representations.
\end{enumerate}
Details of calculations can be found
in~\cite{Almalki19a}, we only indicate milestones here.
The Hamiltonian-independent solution to the analyticity
condition~\eqref{analyticity operator airy} is:
 \begin{equation}
   \label{solution of the analiticity conditon airy function}
   \begin{split}
     \textstyle &f(t;x_1,x_2,x_3)=
     % \exp(\pi\myhbar_4(- E x_{1}^{2}+\frac{\rmi D}{3}x_{1}^{3}))
     \rme^{\pi\myhbar_4(- E x_{1}^{2}+\rmi Dx_{1}^{3}/3)}\,
     % &\times\phi\left(t;Dx_1^{2}+2\rmi Ex_1-2x_3,Dx_1-x_2+\rmi
     %   E\right)
     \phi(t;u_1, u_2)\\
     % \nonumber
     &\text{for }
     % \label{eq:analytic-coordinates}
     u_1=Dx_1^{2}+2\rmi Ex_1-2x_3,\ u_2=Dx_1-x_2+\rmi E.
   \end{split}
 \end{equation}
 The remaining steps will be considered on two specific examples
 satisfying requirements from Prop.~\ref{pr:classification}.

% \subsection{Zero potential case}
% \label{sec:zero-potential-case}

\begin{figure}[htbp]
  \centering
  \includegraphics[scale=.8]{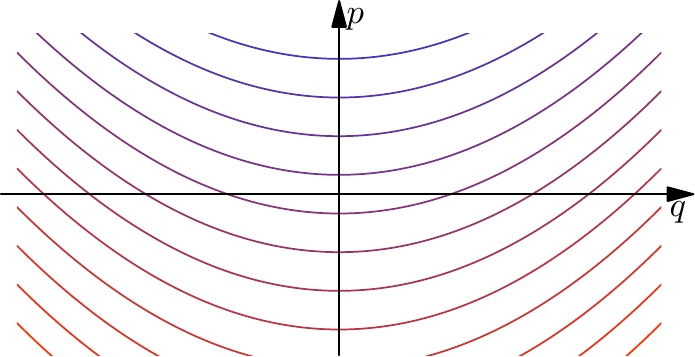}\\[10pt]
  \includegraphics[scale=.8]{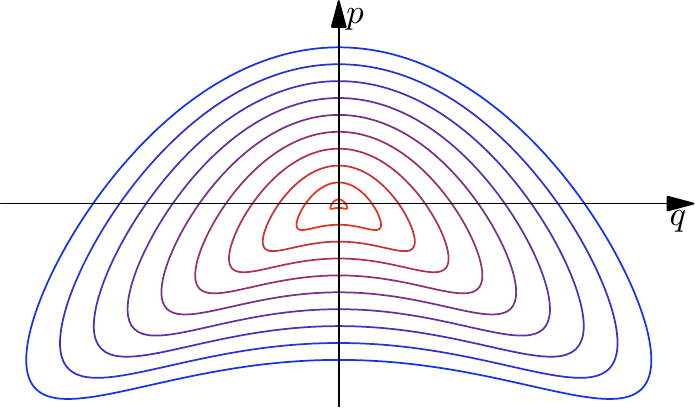}
  \caption{Classical orbits in the phase space of the
    Hamiltonians~\eqref{eq:Hamiltonian-classic-free} (top)
    and~\eqref{eq:classic-Hamiltonian-harmonic} (bottom).}
  \label{fig:class-orbit-phase}
\end{figure}

We consider the matrix satisfying conditions~\eqref{Hamiltonian coefficients}:
\begin{equation}
  \label{specific Hamiltonian coeff free}
  (a_{jk}) =   \frac{1}{m}\begin{pmatrix}
    1 & D  & 0\\
    D   & D^2 &  0\\
    0 & 0 & 0
  \end{pmatrix},
\end{equation}
where \(m\) is the mass.
   The respective Hamiltonian
\begin{equation}
\label{Hamiltonian of quartic anhrmonic}
\begin{split}
  H_{\uir{}{}} % &= \frac{1}{m}
  % (\rmi\rmd\uir{X_1}{})^2+\frac{D^2}{m}(\rmi\rmd\uir{X_2}{})^2+\frac{2D}{m}(\rmi\rmd\uir{X_1}{})(\rmi\rmd\uir{X_2}{})-D\rmd\uir{X_3}{}\\
  & = -\frac{1}{m} (\rmd\uir{X_1}{} +
  D\rmd\uir{X_2}{})^2\,,
\end{split}
\end{equation}
is the Weyl (symmetric) quantisation~\cite{Kisil02e,Kisil10a,Kisil17a} of the classical Hamiltonian
\begin{equation}
  \label{eq:Hamiltonian-classic-free}
  \textstyle  H=\frac{1}{m}(p+Dq^2)^2\,.
\end{equation}
The classical orbits in the phase space are presented on
Fig.~\ref{fig:class-orbit-phase} (top).

The Shr\"odinger equation for adjusted \(H_r\)~\eqref{reduced Hamiltonian equation} with
substitution of~\eqref{reduction polynomials} is a rather complicated first order PDE
on \(\Space{R}{3}\).
However, it significantly simplifies in analytic coordinates~\eqref{solution of the analiticity conditon airy function}:
\begin{equation}
\label{reduced Schrodinger eq anharmonic free}
\begin{split}
 \rmi\myhbar_4\partial_{t}+ H_r&=  \rmi\myhbar_4\partial_{t}+\frac{1}{m}(
  4\pi\rmi\myhbar_4 u_1u_2\frac{\partial}{\partial u_1}
  +4\pi\rmi\myhbar_4 u_2^{2}\frac{\partial}{\partial u_2}\\
  &\quad \qquad +(8\pi^2\myh_2\myhbar_4 u_2^{2}+2\pi\rmi\myhbar_4
  u_2-\pi^2\myhbar_4^{2} u_1^{2})I).
\end{split}
\end{equation}
% where
% \begin{equation}
% u_1=Dx_1^{2}+2\rmi Ex_1-2x_3,\qquad u_2=Dx_1-x_2+\rmi E.
% \end{equation}
The method of characteristics solves \eqref{reduced
  Schrodinger eq anharmonic free} to
\begin{equation}
\label{solution of reduced equation free}
\begin{split}
  \phi(t;u_1,u_2)&=\frac{1}{\sqrt{u_2}}\exp\left(2\pi\rmi\myh_2u_2-\frac{\pi\rmi\myhbar_4}{4}\frac{u_1^{2}}{u_2}\right)\,\\
  &\quad\times \psi \left(-\frac{4\pi}{m}
    t+\frac{1}{u_2},\frac{u_1}{u_2}\right).
\end{split}
\end{equation}
Finally, the structural condition~\eqref{eq:general-annih} turns into
the following Schr\"odinger equation of a free particle:
\begin{equation}
\label{structural equation free}\partial_{\eta\eta}^2\psi(\xi,\eta)+\pi \rmi\myhbar_4\partial_{\xi}\psi(\xi,\eta)=0,
 \end{equation}
for  \(\xi=\frac{1}{u_2}\),  \(\eta=\frac{u_1}{u_2}\). 
%  \begin{equation}
%  \label{xi and eta}
%  \xi=-\frac{4\pi}{m}t+\frac{1}{u_2},\qquad \eta=\frac{u_1}{u_2}. 
% \end{equation}
A generic solution of \eqref{structural equation free} is:
\begin{equation}
  \label{solution of structural equation free}
  \psi(\xi,\eta)=\int_{\mathbb{R}}\rmd s\, g(s)\rme^{-\frac{4\pi\rmi}{\myhbar_4}s^2\xi-2\pi\rmi
    s
    \eta }\,,
\end{equation}
for \(g(s)\) determined by the initial conditions.
Thus, the function \(f\)~\eqref{solution of the analiticity
  conditon airy function} with substitution of \(\phi\) \eqref{solution of reduced
  equation free} and \(\psi\) \eqref{solution of structural equation free}% for
                                % \(\xi\) and \(\eta\) given by
                                % \eqref{xi and eta}
, represents the dynamics of~\eqref{Hamiltonian of quartic anhrmonic}.

% \subsection{Harmonic potential}
% \label{sec:harmonic-potential}

We can similarly treat the Hamiltonian 
%\begin{equation}
%  \label{eq:magnetic-real-line}
%  H_{\myhbar_2\myhbar_4} = \frac{1}{m}\left( \frac{\rmd\ }{\rmd y} +D y^2\right)^2+\frac{a^2}{m}y^2\,,
%\end{equation}
\begin{equation}
  \label{eq:classic-Hamiltonian-harmonic}
    \textstyle  H=\frac{1}{m}(p+Dq^2)^2+\frac{a^2}{m} q^2\,.
\end{equation}
which is different from~\eqref{eq:Hamiltonian-classic-free} by a
quadratic potential with \(a=m\omega > 0\). The classical orbits of
this Hamiltonian are presented on
Fig.~\ref{fig:class-orbit-phase} (bottom).

The analytic coordinates~\eqref{solution of the analiticity
  conditon airy function} simplify the respective first order PDE for~\eqref{reduced
  Hamiltonian equation}; the
 generic solution is: 
\begin{align}
\nonumber
\phi(t;u_1,u_2)&=\frac{\rme^{\frac{2a\pi\rmi}{m}t}}{\sqrt{u_2-\rmi a}}\exp\left(2\pi\rmi\myh_2u_2-\frac{\pi\rmi\myhbar_4}{4}\frac{u_1^{2}}{u_2-\rmi a}\right)\\
\label{solution of reduced equation}
&\times\psi\left(\rme^{\frac{4a\pi\rmi}{m}t}\frac{u_1}{u_2-\rmi a},\rme^{\frac{8a\pi\rmi}{m}t}\frac{u_2+\rmi a}{u_2-\rmi a}\right).
\end{align}

The structural condition \eqref{eq:general-annih} reduces to a heat-like equation:  
\begin{equation}
  \label{heat like eq for structral con}
  \partial_\eta \psi(\xi,\eta)=\frac{1}{2\pi\myhbar_4 a}\partial_{\xi\xi}^2\psi(\xi,\eta).
\end{equation}
A generic solution to~\eqref{heat like eq for structral con} is given by the integral:
\begin{equation}
  \label{solution of heat like for structral con}
  \psi(\xi,\eta)=\left(\frac{a\myhbar_4}{2\eta}
  \right)^{1/2}\int_{\Space{R}{}} \rmd s  \, k(s)\,\rme^{\frac{1}{2}\pi\myhbar_4a\frac{(\xi-s)^2}{\eta}}, 
\end{equation}
where \(k\) is determined by the initial value. 
%  the initial condition \(k(\xi)=g(\xi,0)\) and we use analytic extension from the real variable \(\eta\) to some
%  neighbourhood of the origin in the complex plane.
Thus, the solution \(f\) in form~\eqref{solution of the analiticity conditon airy function} is obtained by substitution of
\(\phi\)~\eqref{solution of reduced equation} and
\(\psi\)~\eqref{solution of heat like for structral con}. 

\section{Discussion and conclusion}
\label{sec:disc-concl}
Hamilton equations describe classical dynamics through a flow on the
phase space. This geometrical picture inspires numerous works
searching for a similar description of quantum evolution starting from
the symplectic structure~\cite{Kibble79a}, curved space-time~\cites{%
  Prugovecki82a,%
  BrookePrugovecki85a,%
  Zachos02a,%
  KaramatskouKleinert14a,%
  Tavernelli16a,%
  HerranzDeLucasTobolski17a}, differential geometry~\cites{
  HurleyVandyck09a,%
  CarienaClemente-GallardoMarmo07a,%
  Clemente-GallardoMarmo08a} and quan\-tilde\-zer--delta\-quan\-tilde\-zer
formalism~\cites{Zachos02a,CiagliaDiCosmoIbortMarmo17a}, coherent states dynamics~\cite{ShalashilinBurghardt08}. A common objective of
those researches is a conceptual similarity between fundamental
geometric objects and their analytical counterparts, e.g. the
symplectic structure on the phase space and derivations of operator
algebras.

This letter is focused on more practical aims. We use an appropriate
coherent state transform (CST) to reduce the order of the
Schr\"odinger equation as described in
Defn.~\ref{de:geometrizing-CS}. As specified in Defn.~\ref{de:method-order-reduction} the application of our method is completely determined by
a choice of a group \(G\), its subgroup \(H\), a representation
\(\uir{}{}\) of \(G\) in a vector space \(V\) and a fiducial vector
\(\ket{0}\in V\). Once these elements are fixed one can give a
characterisation of Hamiltonians admitting geometrisation. For those
systems explicit solutions can be obtained through standard procedures.
Mathematically our work is close to the framework of transmutations of
PDE~\cites{Hersh75a,Carroll85a,KravchenkoTorba16a,%
  KatrakhovSitnik18a,SitnikShishkina19a}.

In Prop.~\ref{pr:classification}, we described all Hamiltonians which
can be geometrized by the minimal nilpotent extension
\(\Space{G}{}\)~\eqref{anh comutation relation} of the Heisenberg
group and the most general fiducial vector
\(\ket{0}\)~\eqref{mother-wavelet for kirilov re} annihilated by the
derived representation of the Lie
algebra~\eqref{eq:derived-representation-first}. This illustrates,
that geometrisation of quantum mechanics may require a set which is
significantly bigger than the classical phase space, despite of
common anticipations~\cite{Zachos02a}.

Our solutions manifest an advantage of the bigger group
\(\Space{G}{}\) over the Heisenberg group even in the simplified case
\(D=0\). The Hamiltonian~\eqref{Hamiltonian of quartic anhrmonic} with
\(D=0\) describes a free quantum particle. Its solution is geometric
in the well-known plane wave decomposition, which emerges
from~\eqref{solution of structural equation free} if we set \(D=E=0\)
in analytic coordinates \((u_1,u_2)\)~\eqref{solution of the
  analiticity conditon airy function}. However, canonical CS do not
provide a transparent solution of a free particle through squeezed
states with \(E\neq 0\).

Similarly, for \(D=0\) the
Hamiltonian~\eqref{eq:classic-Hamiltonian-harmonic} reduces to the
harmonic oscillator. Its geometric dynamic in terms of canonical CS is
only possible for the particular value \(E=m \omega\) of the squeezing
parameter. Yet the larger group \(\Space{G}{}\) allows to obtain a
geometric dynamic for a range of \(E\) as shown
in~\cite{AlmalkiKisil18a}, which also can be recovered from
\(\ket{0}\)~\eqref{solution of reduced equation} for \(D=0\)
in~\eqref{solution of the analiticity conditon airy function}.

Hamiltonians~\eqref{eq:Hamiltonian-classic-free}
and~\eqref{eq:classic-Hamiltonian-harmonic} are similar to charged
particle in a magnetic field. The Fourier transform, which swaps the
coordinate and momentum pictures, relates
\(\ket{0}\)~\eqref{mother-wavelet for kirilov re} to Airy wave
packets~\cites{BerryBalazs79a,ATorre09a} which geometrizes the dual
Hamiltonian:
\begin{equation}
  \label{eq:classic-Hamiltonian-harmonic-dual}
  \textstyle  H%=\frac{1}{m} p^2 + \frac{a^2}{m}(q+Dp^2)^2
  =D^2p^4+\frac{1}{m}(1+2a^2D) q p^2+\frac{a^2}{m}q^2\,.
\end{equation}

Quantisations of Hamiltonians~\eqref{eq:Hamiltonian-classic-free},
\eqref{eq:classic-Hamiltonian-harmonic}
and~\eqref{eq:classic-Hamiltonian-harmonic-dual} may be relevant for
paraxial optics~\cites{ATorre09a,ATorre14a}. The cubic parameter \(D\)
of the fiducial vector \(\ket{0}\)~\eqref{mother-wavelet for kirilov re}
is dictated by the Hamiltonian, while the squeezing parameter \(E\) is
not fixed. However, the convergence of integrals~\eqref{solution of
  heat like for structral con} requires that \(E\geq a=m\omega \).

The research can be continued in many directions, e.g.:
\begin{itemize}
\item Keeping the present group \(\Space{G}{}\) and fiducial
  vector~\(\ket{0}\)~\eqref{mother-wavelet for kirilov re} one may look
  for Hamiltonians beyond the quadratic forms~\eqref{general
    Hamiltonian on g}.
\item Keeping the group \(\Space{G}{}\) look for another fiducial
  vectors, which will be null-solutions to more complicated analytic
  conditions than~\eqref{eq:derived-representation-first}.
\item Finally, many different groups can be considered instead of
  \(\Space{G}{}\) with the Schr\"odinger group~\cite{Niederer72a,%
  KalninsMiller74a,%
  ATorre09a,ATorre14a,%
  Niederer73a,%
  Wolf76a,%
  AldayaGuerrero01a,%
  AldayaCossioGuerreroLopez-Ruiz11b%
} to be
  a very attractive choice.
\end{itemize}
Despite of pragmatic nature of this letter, our method and obtained
results offer a deeper view on correspondence between classical and
quantum mechanics. We use geometrisation, which is very close to that
proposed in~\cite{CiagliaDiCosmoIbortMarmo17a}. However, our
Defn.~\ref{de:geometrizing-CS} has notable distinctions, e.g. it is
compatible with important requirement of positive energies of ground
states.

Although we were not focused on the broader context of geometrisation
of quantum mechanics~\cites{%
  Kibble79a,%,
  Prugovecki82a,%
  BrookePrugovecki85a,%
  Zachos02a,%
  CarienaClemente-GallardoMarmo07a,%
  KaramatskouKleinert14a,%
  Tavernelli16a,%
  HerranzDeLucasTobolski17a,%
  CiagliaDiCosmoIbortMarmo17a,%
  HurleyVandyck09a,Clemente-GallardoMarmo08a}, this aspect is implicitly
present in this work. For example, the symplectic structure flashes
through the obtained solutions~\eqref{solution of reduced equation
  free} and~\eqref{solution of reduced equation}. Indeed, initially
our CST~\eqref{eq:wavelet-transform-shear-group} is parametrised by
points \((x_1,x_2,x_3)\in\Space{R}{3}\) and this odd-dimensional
manifold cannot have a non-degenerate symplectic structure. However,
solutions~\eqref{solution of reduced equation free}
and~\eqref{solution of reduced equation} are defined in terms of
function \(\psi\) of two complex variables. The complex structure in
the first variable is compatible with the symplectic structure on the
classical phase space with coordinate \((x_1,x_3)\). On the other
hand, the complex structure in the second variable requires an
analytic extensions, see discussion of this in~\cite{AlmalkiKisil18a}.
\begin{figure}%[htbp]
  %\centering
  \includegraphics{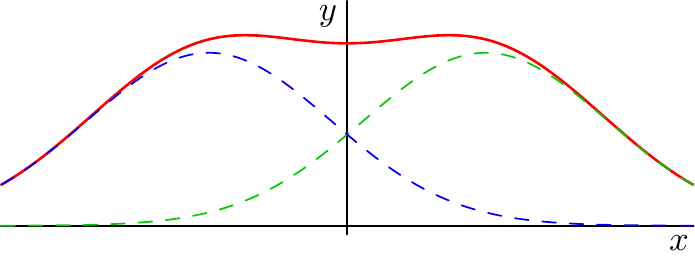}\\[.5em]
  \includegraphics{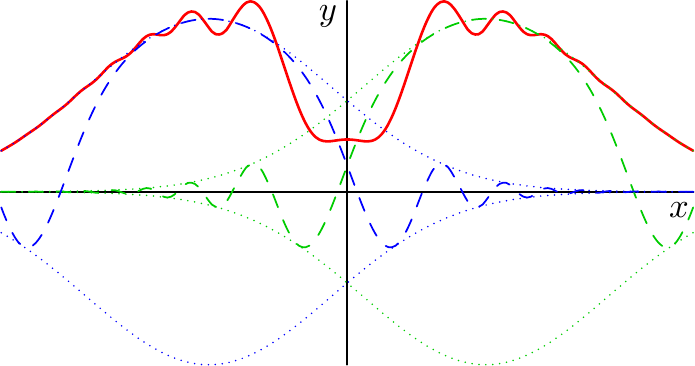}\\
  \caption[Interference of two displaced ground states]
  {Interference of two displaced ground states.%\par
    For \(D=0\) the top
    graph shows interference of two Gaussians (drawn in dashed blue
    and green lines). The magnitude of interference (drawn in
    solid red) is simply the sum of the respective magnitudes since 
    phases of ground states coincide.%\par
    For \(D\neq 0\) the bottom graph shows the interference (drawn in
    solid red) of two displaced ground states~\eqref{mother-wavelet
      for kirilov re}. The real parts of the ground states are drawn
    in dashed blue and green lines, envelopes of their magnitudes are
    drawn by dotted lines of the respective colours,
    cf. Fig.~\ref{fig:graund-state-harm}. There is the clear
    interference pattern since phases of the ground states are
    oscillating with variable frequencies.}
  \label{fig:interference}
\end{figure}
Another prominent theme of geometrisation is a correspondence between
classical and quantum mechanics.
Hamiltonians~\eqref{eq:Hamiltonian-classic-free}
and~\eqref{eq:classic-Hamiltonian-harmonic} resemble particles in
magnetic field with velocity-dependent forces, which do not produce a
work.  Classical dynamics determined
by~\eqref{eq:Hamiltonian-classic-free}
and~\eqref{eq:classic-Hamiltonian-harmonic} in the configuration space
is independent of \(D\).  with \(D=0\) (no field).  However, classical
trajectories in the phase space for \(D\neq 0\) are significantly
different from the rigid rotation of the phase space familiar from the
harmonic oscillator, see Fig.~\ref{fig:class-orbit-phase}
(bottom). Correspondingly, the quantum ground state
\(\ket{0}\)~\eqref{mother-wavelet for kirilov re} in the coordinate
representation has the same density \(\modulus{\ket{0}}\) for every \(D\)
and it coincides with the Gaussian, see
Fig.~\ref{fig:graund-state-harm}. However, quantum phase factors of
the ground state \(\ket{0}\)~\eqref{mother-wavelet for kirilov re}
depends on \(D\) and it can be visualised by interference of two
ground states slightly displaced in space, see
Fig.~\ref{fig:interference}. This again illustrates the fundamental
correspondence between classical momenta and quantum phases.

At a wider scope, the link between classical and quantum mechanics is
a two-way road: geometrisation of quantum mechanics (discussed in this
letter) is naturally complemented by ``non-commutativisation'' (as a
form of quantisation) of classical theory. Indeed, since Dirac's
paper~\cite{Dirac26b} non-commutativity is deemed to be the main
distinguishing assumption of quantum mechanics from the classical
theory. The degree of non-commutativity is measured by a non-zero
Planck constant \(\myhbar\) and the correspondence to classical
mechanic is implemented through the semi-classical limit \(\myhbar \rightarrow
0\). It was only recently shown~\cites{Kisil12c,Kisil17a} that there
is a natural formulation of classical mechanics with non-commutative
observables and a non-zero Planck constant. This quantum-like form of classical mechanics was based on the Heisenberg group
representations, which is linked it to the present work. Further research
in this direction seems to be promising.

\medskip

\noindent\textbf{Acknowledgements:} We are grateful to Prof.~D.\,Shalashilin and anonymous referees for several useful suggestions.

%\bibliography{abbrevmr,akisil,analyse,algebra,arare,aclifford,aphysics,acompute,ageometry,acombin}
\providecommand{\noopsort}[1]{} \providecommand{\printfirst}[2]{#1}
  \providecommand{\singleletter}[1]{#1} \providecommand{\switchargs}[2]{#2#1}
  \providecommand{\irm}{\textup{I}} \providecommand{\iirm}{\textup{II}}
  \providecommand{\vrm}{\textup{V}} \providecommand{\cprime}{'}
  \providecommand{\eprint}[2]{\texttt{#2}}
  \providecommand{\myeprint}[2]{\texttt{#2}}
  \providecommand{\arXiv}[1]{\myeprint{http://arXiv.org/abs/#1}{arXiv:#1}}
  \providecommand{\doi}[1]{\href{http://dx.doi.org/#1}{doi:
  #1}}\providecommand{\CPP}{\texttt{C++}}
  \providecommand{\NoWEB}{\texttt{noweb}}
  \providecommand{\MetaPost}{\texttt{Meta}\-\texttt{Post}}
  \providecommand{\GiNaC}{\textsf{GiNaC}}
  \providecommand{\pyGiNaC}{\textsf{pyGiNaC}}
  \providecommand{\Asymptote}{\texttt{Asymptote}}

\end{document}